\ifdefined\pdfminorversion\pdfminorversion=6\fi
\documentclass[aps,prd,twocolumn,english,preprintnumbers,amsmath,amssymb,nofootinbib,superscriptaddress]{revtex4-2}

\usepackage{hyperref}
\usepackage{graphicx}
\usepackage{xcolor}
\usepackage{amsmath}
\usepackage{balance}

\graphicspath{{figures-N=2/}}

\newcommand{\beq}{\begin{equation}}
\newcommand{\eeq}{\end{equation}}
\newcommand{\bqa}{\begin{eqnarray}}
\newcommand{\eqa}{\end{eqnarray}}

\def\sumint{\hbox{$\sum$}\!\!\!\!\!\!\!\int}
\def\square{\vcenter{\vbox{\hrule height.4pt
          \hbox{\vrule width.4pt height4pt
          \kern4pt\vrule width.3pt}\hrule height.4pt}}}

\newcommand{\symnt}{\text{SYM}_{2,4}}
\newcommand{\symos}{\text{SYM}_{1,6}}

\newcommand{\symot}{\text{SYM}_{1,10}}
\def\symod{\text{SYM}_{1,{\cal D}}}
\newcommand{\Nc}{N_c}

\newcommand{\dA}{d_A}
\newcommand{\mub}{\bar\mu}

\newcommand{\Lb}{\log{\mub\over4\pi T}}
\newcommand{\Zone}{{\zeta'(-1)\over\zeta(-1)}}
\newcommand{\Zthree}{{\zeta'(-3)\over\zeta(-3)}}
\newcommand{\trI}{{\rm Tr}\,\mathbf{1}}
\newcommand{\x}{\left({\lambda\over\pi^2}\right)}

\begin{document}
\raggedbottom

\title{\texorpdfstring{${\cal N}=2$}{N=2} supersymmetric Yang-Mills thermodynamics from effective field theory}

\author{Ubaid Tantary}
\email{utantary@pmu.edu.sa}
\affiliation{Department of Mathematics and Natural Sciences,
Prince Mohammad Bin Fahd University, Al Khobar 31952, Saudi Arabia}

\author{Qianqian Du}
\email{dqianqian@mailbox.gxnu.edu.cn}
\affiliation{Department of Physics, Guangxi Normal University,
Guilin 541004, China}
\affiliation{Guangxi Key Laboratory of Nuclear Physics and Technology,
Guilin 541004, China}

\date{\today}

\begin{abstract}
We compute the weak-coupling free energy density of pure four-dimensional
${\cal N}=2$ supersymmetric Yang--Mills (SYM) theory through second order in
the 't Hooft coupling $\lambda$ at finite temperature using
effective-field-theory methods.  The contribution from the hard scale $T$
is obtained from massless three-loop vacuum diagrams in the full theory,
and that from the electric scale $\sqrt{\lambda}\,T$ from a two-loop
calculation in the dimensionally reduced three-dimensional effective
theory.  In contrast to ${\cal N}=4$ SYM, the theory is asymptotically
free: the coupling-renormalization counterterm, together with the EFT
unit-operator counterterm, cancels all $1/\epsilon$ poles.  The remaining
renormalization-scale dependence is fixed by the one-loop beta function.
The expansion reproduces earlier results through order
$\lambda^{3/2}$, and the order-$\lambda^2$ term is new.
\end{abstract}

\keywords{Finite-temperature field theory, Thermodynamics, Supersymmetric field theory, Dimensional reduction}

\maketitle

\section{Introduction}

Supersymmetric Yang-Mills (SYM) theories provide theoretically controlled
examples of thermal field theories.  At weak coupling, their perturbative finite-temperature
expansions exhibit the same separation of scales that occurs in hot QCD:
the hard scale $T$, the electric scale $gT$, and the magnetic scale
$g^2T$, where $g$ is the four-dimensional gauge
coupling~\cite{Kapusta,Gross:1980br,Linde:1980ts,Arnold:1994ps,Arnold:1994eb,Braaten:1995jr}.
At the same time, supersymmetry strongly constrains the field content and
renormalization structure.  The maximally supersymmetric
four-dimensional theory, ${\cal N}=4$ SYM, has been studied extensively at
finite temperature.  Its weak-coupling free energy is known through order
$\lambda^2$~\cite{Du:2021pqa} and has also been derived with
effective-field-theory methods~\cite{Andersen:2022N4EFT}.  The next
perturbative term, of order $\lambda^{5/2}$, was recently computed in
Ref.~\cite{Carrington:2026N4lambda52}.  In the large-$\Nc$ limit, its
strong-coupling thermodynamics can be studied using
holography~\cite{Maldacena:1997re,Gubser:1998nz}.

Two natural extensions of the four-dimensional ${\cal N}=4$ calculation are
possible.  One can keep maximal supersymmetry and change the spacetime
dimension by reducing ten-dimensional ${\cal N}=1$ SYM to dimensions other
than four, or one can keep four spacetime dimensions and reduce the number
of supercharges.  We consider the latter and study pure ${\cal N}=2$ SYM,
the ${\cal N}=2$ analogue of pure Yang-Mills theory: the only fields are
the gauge field, two adjoint Majorana fermions, and two real adjoint
scalars.  This extends the ${\cal N}=4$ EFT calculation to an
asymptotically free theory with eight supercharges.

The thermodynamics of pure $SU(2)$ ${\cal N}=2$ SYM was studied by
Paik and Yaffe~\cite{Paik:2009iz}, who analyzed both the low-temperature
Seiberg-Witten regime and the asymptotically high-temperature plasma.  In
the high-temperature limit, they gave the
equilibrium free energy through order $g^3$.  Earlier, Vazquez-Mozo
computed the two-loop free energy for supersymmetric Yang-Mills theories
with various numbers of supercharges and the leading nonanalytic ring
contribution in four dimensions~\cite{VazquezMozo:1999ge}.  Translated
into the present notation, these results give the ${\cal N}=2$ free
energy through order $\lambda^{3/2}$.

Here we compute the next term, the finite order-$\lambda^2$ correction.
This requires the hard three-loop vacuum contribution, the soft
two-loop contribution, the EFT unit-operator counterterm, and the
coupling-renormalization counterterm required by the running of
the ${\cal N}=2$ gauge coupling.  This last contribution is absent in
the conformal ${\cal N}=4$ calculation, where the beta function vanishes.
The result is finite, and its explicit renormalization-scale dependence is
fixed by the one-loop beta function.

As in the ${\cal N}=4$ EFT calculation~\cite{Andersen:2022N4EFT}, we
combine two reductions.  SUSY dimensional reduction from $\symos$ fixes the
four-dimensional field content.  High-temperature dimensional reduction
onto a three-dimensional effective theory separates the hard and soft
thermal
scales~\cite{Ginsparg:1980ef,Appelquist:1981vg,Kajantie:1995dw,Braaten:1995cm}.
The hard contribution is obtained from the same massless parent-theory
vacuum diagrams, evaluated for six-dimensional ${\cal N}=1$ SYM; the EFT
contains two real adjoint scalars and the corresponding screening masses.

Our paper is organized as follows.  Section~\ref{sec:susy} specifies the
four-dimensional theory and its six-dimensional origin.  The thermal EFT is
constructed in
Sec.~\ref{sec:dimreduction}, and its matching coefficients are obtained in
Sec.~\ref{sec:params}.  Section~\ref{sec:eft} contains the soft calculation and the counterterms.
The result and its checks are given in Sec.~\ref{sec:result}, followed by the
summary in Sec.~\ref{sec:conclusions}.  Technical details are collected in
the appendices.

{\em Notation:} We write
\beq
\begin{aligned}
  {\rm SYM}_{2,4}&\equiv
  \text{four-dimensional pure }{\cal N}=2\text{ SYM},
  \\
  {\rm SYM}_{1,6}&\equiv
  \text{six-dimensional }{\cal N}=1\text{ SYM}.
\end{aligned}
\eeq
In what follows we abbreviate these theories as $\symnt$ and $\symos$,
respectively.  The first label is the conventional ${\cal N}$ of the theory
in that spacetime dimension, and the second label is the spacetime dimension.
Thus ${\rm SYM}_{1,6}$ and ${\rm SYM}_{2,4}$ both have eight supercharges and
are related by SUSY dimensional reduction.  We use the
mostly-minus metric convention, lower-case letters for Minkowski momenta,
and upper-case letters for Euclidean momenta.  The renormalization scale is
denoted by $\bar\mu$.  In the full thermal theory,
$P=(P_0,{\bf p})$ with bosonic Matsubara frequencies $P_0=2\pi nT$ and
fermionic frequencies $P_0=(2n+1)\pi T$.  In the dimensionally reduced
effective theory, the fields are static bosonic zero modes.  We use
\beq
  \lambda=g^2 C_A .
\eeq
The generators are normalized by
$\mathrm{Tr}(T^aT^b)=\delta^{ab}/2$, with
$[T^a,T^b]=if^{abc}T^c$,
$f^{acd}f^{bcd}=C_A\delta^{ab}$.  Here $C_A$ is the adjoint quadratic
Casimir, and $d_A=\delta^{aa}$ is the dimension of the adjoint
representation.
For $SU(\Nc)$, $C_A=\Nc$ and $d_A=\Nc^2-1$.

\section{Supersymmetric Yang-Mills theory}
\label{sec:susy}

We start with the ${\cal N}=1$ supersymmetric Yang-Mills Lagrangian in
${\cal D}$-dimensional Minkowski space and general covariant
gauge~\cite{Brink:1976bc,VazquezMozo:1999ge}
\beq \label{lagN1N2}
\begin{aligned}
{\cal L}_{\symod}
={\rm Tr}\biggl[
&-{1\over2}G_{MN}G^{MN}
+2i\bar\psi\Gamma^M D_M\psi
\\
&-{1\over\xi}(\partial^M A_M)^2
-2\bar\eta\,\partial^M D_M\eta
\biggr].
\end{aligned}
\eeq
Here $M,N=0,\ldots,{\cal D}-1$ and
$G_{MN}=\partial_MA_N-\partial_NA_M-ig[A_M,A_N]$.  The adjoint covariant
derivative is $D_M=\partial_M-ig[A_M,\cdot]$, with $g$ the gauge coupling.
The $\Gamma^M$ are Dirac matrices in ${\cal D}$ dimensions, and $\psi$ is
an adjoint fermion.
The gauge group is $SU(\Nc)$, with $\Nc$ the number of colors.  The fields
$\eta$ and $\bar\eta$ are the ghost and antighost fields, respectively, and
$\xi$ is the gauge parameter.

The maximal spacetime dimensions admitting SYM theories with
$n_{\rm SC}=16,8,$ and $4$ supercharges are
\bqa
n_{\rm SC}=16&\rightarrow&{\cal D}_{\rm max}=10\;,\nonumber\\
n_{\rm SC}=8&\rightarrow&{\cal D}_{\rm max}=6\;,\nonumber\\
n_{\rm SC}=4&\rightarrow&{\cal D}_{\rm max}=4\;.
\label{eq:nscN2}
\eqa
The fermion obeys a Majorana-Weyl condition in ten dimensions and a Weyl
condition in six and four dimensions; in four dimensions the Weyl fermions
may equivalently be written as Majorana spinors.  Equality
of the bosonic and fermionic on-shell degrees of freedom fixes the spinor trace,
$\trI={\cal D}_{\rm max}-2$.

The theories of interest are obtained by reducing Eq.~\eqref{lagN1N2} to
an integer number of spacetime dimensions $D\leq{\cal D}_{\rm max}$,
keeping all $n_{\rm SC}$ supercharges.  To preserve
supersymmetry, we use the regularization by dimensional reduction (RDR)
scheme~\cite{Siegel:1979wq,Capper:1979ns,Avdeev:1982xy,Stockinger:2005gx}.
In this scheme, the fields in Eq.~\eqref{lagN1N2} remain
${\cal D}$-dimensional tensors or
spinors, while loop momenta are continued to $d=D-2\epsilon$
dimensions~\cite{VazquezMozo:1999ge}.

The $\symnt$ theory is obtained by SUSY dimensional reduction of $\symos$
with ${\cal D}={\cal D}_{\rm max}=6$.  The gauge group remains $SU(\Nc)$,
and all fields are in the adjoint representation.  Splitting the
six-dimensional index as
$M=(\mu,m)$, with $\mu=0,1,2,3$ and $m=4,5$, we write the gauge field as
$A_M=(A_\mu,A_m)$.  The two internal components $A_4$ and $A_5$ transform as
scalars under the four-dimensional Lorentz group and are denoted by
$\Phi_I=(X,Y)$, $I=1,2$, where $X$ and $Y$ are Hermitian scalar and
pseudoscalar fields, respectively.  The six-dimensional spinor gives two
adjoint four-dimensional Majorana fermions $\psi_i$, $i=1,2$.  The bosonic
and fermionic on-shell degrees of freedom are both four, and $\trI=4$.
Paik and Yaffe use one complex adjoint scalar and two adjoint Weyl
fermions~\cite{Paik:2009iz}.  For the calculation we use the equivalent
real-scalar and Majorana-fermion notation of
Ref.~\cite{Andersen:2022N4EFT}.  In these conventions, the Minkowski-space
Lagrangian is
\beq
\begin{aligned}
\mathcal L_{\symnt}
={}&{\rm Tr}\biggl[
 -{1\over2}G_{\mu\nu}G^{\mu\nu}
 +(D_\mu\Phi_I)(D^\mu\Phi_I)
 +i\bar\psi_i\,{\not\!\! D}\,\psi_i
\\
&\hspace{1.0cm}
 -{1\over2}g^2\bigl(i[\Phi_I,\Phi_J]\bigr)^2
\\
&\hspace{1.0cm}
 -ig\,\bar\psi_i
 \bigl[\alpha_{ij}X+i\beta_{ij}\gamma_5Y,\psi_j\bigr]
\biggr]
\\
&+\mathcal L_{\rm gf}^{\symnt}
+\mathcal L_{\rm gh}^{\symnt}
+\Delta\mathcal L_{\symnt},
\end{aligned}
\label{lagN2}
\eeq
where $G_{\mu\nu}=\partial_\mu A_\nu-\partial_\nu A_\mu
-ig[A_\mu,A_\nu]$ and $D_\mu=\partial_\mu-ig[A_\mu,\cdot]$.  The
$2\times2$ matrices $\alpha$ and $\beta$ act on the two Majorana fermions.
Restricting the matrices of Ref.~\cite{Du:2021pqa} to the ${\cal N}=2$
field content gives the convenient representation
\[
  \alpha=i\sigma_2,\qquad \beta=-i\sigma_2 ,
\]
where $\sigma_2$ is the second Pauli matrix.  For the Majorana Yukawa
interaction in Eq.~\eqref{lagN2}, only the antisymmetric parts of
$\alpha$ and $\beta$ contribute.  Together with
$\alpha^2=\beta^2=-\mathbf 1_2$, this fixes
$\alpha=\pm i\sigma_2$ and $\beta=\pm i\sigma_2$.  The signs can be
absorbed into redefinitions of $X$ and $Y$, and $[\alpha,\beta]=0$
follows.
Equation~\eqref{lagN2} has the same form as the ${\cal N}=4$
Lagrangian because both are obtained from the $\symod$ Lagrangian by SUSY
dimensional reduction.  For ${\cal N}=2$, the scalar and Majorana-fermion
index ranges are $I=1,2$ and $i=1,2$, rather than $A=1,\ldots,6$ and
$i=1,\ldots,4$.  All differences from the ${\cal N}=4$ calculation enter
through these multiplicities, the spinor trace, and the parent
dimension ${\cal D}_{\rm max}$.

For the four-dimensional theory we work in general covariant gauge, giving
\bqa
\mathcal L_{\rm gf}^{\symnt}
&=&-\frac{1}{\xi}{\rm Tr}\bigl[(\partial^\mu A_\mu)^2\bigr],
\\
\mathcal L_{\rm gh}^{\symnt}
&=&-2\,{\rm Tr}\bigl[\bar\eta\,\partial^\mu D_\mu\eta\bigr].
\eqa
Finally, the term $\Delta\mathcal L_{\symnt}$ contains the ultraviolet
counterterms.

\section{Dimensional reduction at finite temperature}
\label{sec:dimreduction}

At high temperature, the nonzero Matsubara modes act as heavy fields with
masses of order $T$, while interactions with the plasma generate screening
masses of order $gT$ for the temporal gauge field and the adjoint scalars.
At weak coupling these scales are well separated, and static correlation
functions can be reproduced by a three-dimensional effective theory for the
bosonic zero modes, with parameters that encode the physics of the hard
scale
$T$~\cite{Ginsparg:1980ef,Appelquist:1981vg,Kajantie:1995dw,Braaten:1995cm,Braaten:1995jr}.
Static magnetic fields are not screened at any finite order of perturbation
theory; the magnetic scale $g^2T$ must be treated nonperturbatively and, as
in QCD, first affects the free energy at order
$\lambda^3$~\cite{Linde:1980ts,Gross:1980br}, beyond the order considered
here.  We work in the high-temperature phase, where the scalar fields
have vanishing expectation values~\cite{Paik:2009iz}.

The thermal partition function of the full theory is
\beq
\begin{aligned}
{\cal Z}_{\symnt}
={}&
\int{\cal D}\bar\eta{\cal D}\eta
{\cal D}A_\mu{\cal D}\Phi_I
\\
&\times
{\cal D}\bar\psi_i{\cal D}\psi_i\,
\exp\biggl[
-\int_0^\beta d\tau
\int d^3x\,{\cal L}_{\symnt}
\biggr].
\end{aligned}
\eeq
Here $\beta=1/T$, $\tau$ is Euclidean time, and ${\cal L}_{\symnt}$
denotes the Euclidean continuation of Eq.~\eqref{lagN2}.  The functional
integral runs over the fields of Eq.~\eqref{lagN2}.  The gauge, scalar, and
ghost fields obey periodic
boundary conditions, while the fermions obey anti-periodic boundary
conditions.  The free energy density is
\beq
{\cal F}= -{1\over\beta V}\log{\cal Z}_{\symnt},
\label{FdefN2}
\eeq
where $V$ is the spatial volume.  Here and below, ${\cal F}$ without a
subscript denotes the full free-energy density; individual contributions
are shown explicitly in the numerators of normalized ratios.

After integrating out the non-static modes, the static sector is described
by a three-dimensional effective theory containing the spatial gauge field
$A_i$, $i=1,2,3$, the temporal gauge field $A_0$, and the two real adjoint
scalars $\Phi_I$.  We call this theory electrostatic SYM (ESYM), in
analogy with electrostatic QCD~\cite{Braaten:1995jr}.  There are no fermion
zero modes because the fermions are anti-periodic.  Equating the full and
effective partition functions gives
\beq
\begin{aligned}
{\cal Z}_{\symnt}
={}&
\int{\cal D}\bar\eta{\cal D}\eta
{\cal D}A_i{\cal D}A_0{\cal D}\Phi_I\,
\exp\biggl[
-f_EV
\\
&\hspace{1.2cm}
-\int d^3x\,{\cal L}_{\rm ESYM}^{{\cal N}=2}
\biggr].
\end{aligned}
\label{zesymN2}
\eeq
Here $f_E$ is the coefficient of the unit operator and gives the hard
contribution $Tf_E$ to the four-dimensional free-energy density.  The
non-static modes, including all fermionic modes, contribute to $f_E$ and the
matched EFT parameters.  Denoting the functional integral over the ESYM
fields $A_i$, $A_0$, $\Phi_I$, $\eta$, and $\bar\eta$ by
${\cal Z}_{\rm ESYM}$, the soft contribution is
$f_M=-\log{\cal Z}_{\rm ESYM}/V$, and the full free-energy density is
${\cal F}=T(f_E+f_M)$.  Up to normalization, the EFT fields are the static
modes of the full theory.  At zero chemical potential, the symmetries
$A_0\to-A_0$ and $\Phi_I\to-\Phi_I$ exclude terms odd in these
fields~\cite{Paik:2009iz}.  The terms in the effective Lagrangian required
through order $\lambda^2$ are
\beq
\begin{aligned}
{\cal L}_{\rm ESYM}^{{\cal N}=2}
={}&
{1\over2}{\rm Tr}\,G_{ij}^2
+{\rm Tr}\left[(D_iA_0)(D_iA_0)\right]
\\
&
+{\rm Tr}\left[(D_i\Phi_I)(D_i\Phi_I)\right]
+m_E^2{\rm Tr}\,A_0^2
\\
&
+M^2{\rm Tr}\,\Phi_I^2
+h_E{\rm Tr}\left[(i[A_0,\Phi_I])^2\right]
\\
&
+{1\over2}g_3^2{\rm Tr}\left[(i[\Phi_I,\Phi_J])^2\right]
\\
&+{\cal L}_{\rm gf}^{\rm ESYM}+{\cal L}_{\rm gh}^{\rm ESYM}
+\delta{\cal L}_{\rm ESYM},
\end{aligned}
\label{lagesymN2}
\eeq
where the gauge fields are expanded as $A_i=A_i^aT^a$ and
$A_0=A_0^aT^a$.  The scalar fields are similarly expanded in the basis of
color generators, $\Phi_I=\Phi_I^aT^a$, giving
\bqa
(D_iA_0)^a&=&\partial_iA_0^a+g_Ef^{abc}A_i^bA_0^c,
\\
(D_i\Phi_I)^a&=&\partial_i\Phi_I^a+g_Ef^{abc}A_i^b\Phi_I^c.
\eqa
Here $g_E$ is the gauge coupling of the effective theory, and the
three-dimensional field strength is
$G_{ij}^a=\partial_iA_j^a-\partial_jA_i^a+g_Ef^{abc}A_i^bA_j^c$.
The three-dimensional gauge-fixing and ghost terms are
${\cal L}_{\rm gf}^{\rm ESYM}=-{1\over\xi}{\rm Tr}\bigl[(\partial_iA_i)^2\bigr]$
and
${\cal L}_{\rm gh}^{\rm ESYM}=-2\,{\rm Tr}\bigl[\bar\eta\,\partial_iD_i\eta\bigr]$.
Finally, $\delta{\cal L}_{\rm ESYM}$ collects the higher-order local
operators consistent with gauge and rotational invariance and the
reflection symmetries above.  The quartic $A_0$
self-interactions $({\rm Tr}A_0^2)^2$ and ${\rm Tr}A_0^4$ are examples of
such operators; their first contribution to the free energy arises at
order $\lambda^3$, and they are therefore not displayed separately in
Eq.~\eqref{lagesymN2}.  Equation~\eqref{lagesymN2} is the canonically
normalized form
of the $SU(2)$ ESYM Lagrangian of Paik and Yaffe~\cite{Paik:2009iz}, written
here for $SU(\Nc)$.  Through order $\lambda^2$,
it has the same operator structure as the ${\cal N}=4$ EFT: both static
sectors contain $A_i$, $A_0$, and adjoint scalar zero modes, with gauge and
scalar interactions inherited from the common $\symod$ parent action.  For
${\cal N}=2$, the scalar range is $I=1,2$, rather than $A=1,\ldots,6$; we
write $M^2$ for the scalar mass parameter, corresponding to $m_S^2$ in the
${\cal N}=4$ EFT.  The soft diagrams therefore have the same
topologies, while their multiplicities and the matched parameters differ.

\section{Parameters of the effective theory}
\label{sec:params}

The parameters of the effective theory are fixed by matching the full and
effective theories in strict perturbation
theory~\cite{Braaten:1995cm,Braaten:1995jr}: massless propagators are
used on both sides of the matching, the mass terms of the EFT are treated as
perturbations, and the resulting infrared divergences cancel between the two
sides.  The massive EFT propagators are used only when computing soft
contributions in Sec.~\ref{sec:eft}.

\subsection{Coefficient of the unit operator}
\label{sec:hard}

\begin{figure*}[t]
\begin{center}
\includegraphics[width=0.72\textwidth]{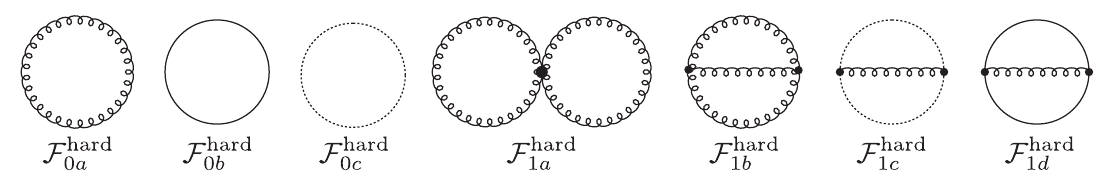}
\end{center}
\vspace{-4mm}
\caption{One- and two-loop vacuum diagrams contributing to the $\symod$
free energy.  For the present calculation these diagrams are
evaluated in $\symos$, with ${\cal D}_{\rm max}=6$ and $\trI=4$.  Spiral
lines denote $\symod$ gauge fields, solid lines denote $\symod$
fermions, and dotted lines denote ghosts.  Reproduced without modification
from Ref.~\cite{Andersen:2022N4EFT}.}
\label{fig:hard12N2}
\end{figure*}

Equating the full and effective partition functions in strict perturbation
theory and taking the logarithm gives
\beq
  f_EV-\log{\cal Z}_{\rm ESYM}=-\log{\cal Z}_{\symnt}.
\eeq
In strict perturbation theory the EFT vacuum diagrams are
scaleless and vanish in dimensional regularization.  Thus $Tf_E$ is the
hard contribution to the four-dimensional free energy density,
\beq
  {\cal F}_{\rm hard}=T f_E.
\eeq
At leading order, this hard contribution is the ideal-gas free energy.  For
${\cal N}=2$ SYM the bosonic and fermionic on-shell degrees of freedom
per adjoint generator are
\beq
  n_b=2+2=4,\qquad n_f=4,
\label{nbnfN2}
\eeq
and hence
\beq
\begin{aligned}
\mathcal F_{\rm ideal}^{{\cal N}=2}
&=
-{\pi^2T^4\over90}\left(n_b+{7\over8}n_f\right)d_A
\\
&=
-{\pi^2T^4\over90}\left(4+{7\over8}\cdot4\right)d_A
=-{d_A\pi^2T^4\over12}.
\end{aligned}
\label{FidealN2}
\eeq
The one- and two-loop graphs in $\symod$ are evaluated with
${\cal D}=6$ and $\trI=4$; together with SUSY dimensional reduction to
$\symnt$, they give the one- and two-loop hard contributions.  The relevant
diagrams are shown in Fig.~\ref{fig:hard12N2}.

The general-${\cal D}$ expressions are those of
Refs.~\cite{Du:2021pqa,Andersen:2022N4EFT}.
The sum of the one-loop graphs ${\cal F}_{0a}^{\rm hard}$,
${\cal F}_{0b}^{\rm hard}$, and ${\cal F}_{0c}^{\rm hard}$ is
\beq
{\cal F}_{0}^{\rm hard}
={1\over2}d_A({\cal D}-2)(f_0'-b_0') .
\label{F0hardSymodN2}
\eeq
For ${\cal D}=6$, this gives Eq.~\eqref{FidealN2}.  The quantities $b_0'$
and $f_0'$ are defined in Eqs.~\eqref{b0primeN2} and~\eqref{f0primeN2}.

The four two-loop graphs in Fig.~\ref{fig:hard12N2} give
\beq
{\cal F}_{1}^{\rm hard}
=d_A\lambda\left[
{\cal F}_{1a}^{\rm hard}
 +{\cal F}_{1b}^{\rm hard}
 +{\cal F}_{1c}^{\rm hard}
 +{\cal F}_{1d}^{\rm hard}
\right],
\eeq
where
\bqa
{\cal F}_{1a}^{\rm hard}
&=&
{{\cal D}({\cal D}-1)\over4}b_1^2,
\\
{\cal F}_{1b}^{\rm hard}
&=&
-{3\over4}({\cal D}-1)b_1^2,
\\
{\cal F}_{1c}^{\rm hard}
&=&
{1\over4}b_1^2,
\\
{\cal F}_{1d}^{\rm hard}
&=&
{({\cal D}-2)\over4}\trI
\left[f_1^2-2f_1b_1\right].
\eqa
Using $\trI={\cal D}-2$, this yields
\beq
{\cal F}_{1}^{\rm hard}
=d_A\lambda{({\cal D}-2)^2\over4}(b_1-f_1)^2 .
\label{F1hardSymodN2}
\eeq
Using the leading part of Eq.~\eqref{b1minusf1N2},
$b_1-f_1=T^2/8+O(\epsilon)$, and setting ${\cal D}=6$ gives
\beq
{\cal F}_{0}^{\rm hard}+{\cal F}_{1}^{\rm hard}
=
-{d_A\pi^2T^4\over12}
\left[
1-{3\over4}{\lambda\over\pi^2}
+O(\lambda^2)
\right].
\eeq
The coupling-renormalization counterterm contains a $1/\epsilon$ pole, so the
normalized order-$\lambda$ hard contribution is required through
$O(\epsilon)$:
\beq
\begin{aligned}
{{\cal F}_{1}^{\rm hard}\over{\cal F}_{\rm ideal}}
={}&
{\lambda\over\pi^2}
\biggl[
-{3\over4}
\;+\;
\epsilon\biggl(
-3\log{\mub\over4\pi T}
-3{\zeta'(-1)\over\zeta(-1)}
\\
&\hspace{2.0cm}
+\log2-3
\biggr)
\biggr].
\end{aligned}
\label{OepslambdaN2}
\eeq

\subsubsection{Three-loop hard term}

\begin{figure*}[t]
\begin{center}
\includegraphics[width=0.92\textwidth]{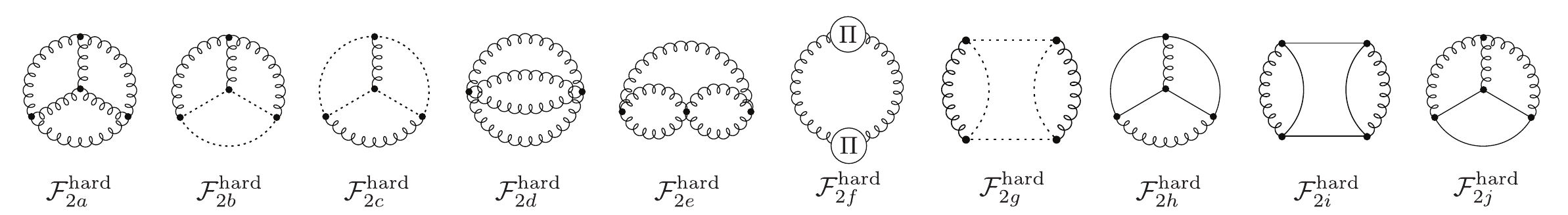}
\end{center}
\vspace{-4mm}
\caption{Three-loop vacuum diagrams contributing to the $\symod$
free energy.  The notation is the same as in Fig.~\ref{fig:hard12N2}.  A
circle labeled $\Pi$ denotes the one-loop $\symod$ gauge-field
self-energy.  Reproduced without modification from
Ref.~\cite{Andersen:2022N4EFT}; these diagrams were first introduced for
$\symot$ in Ref.~\cite{Du:2021pqa}.}
\label{fig:hard3N2}
\end{figure*}

The three-loop vacuum graphs are shown in Fig.~\ref{fig:hard3N2}.  Their
general-${\cal D}$ contributions were obtained in
Refs.~\cite{Du:2021pqa,Andersen:2022N4EFT}.  For the present calculation they
are evaluated at ${\cal D}_{\rm max}=6$ and $\trI=4$.
The general-${\cal D}$ result is
\beq
\begin{aligned}
{\cal F}_2^{\rm hard}
={}&
d_A\lambda^2\bigl[
{\cal F}_{2a}^{\rm hard}
 +{\cal F}_{2b}^{\rm hard}
 +{\cal F}_{2c}^{\rm hard}
 +{\cal F}_{2d}^{\rm hard}
\\
&\quad
 +{\cal F}_{2e}^{\rm hard}
 +{\cal F}_{2f}^{\rm hard}
 +{\cal F}_{2g}^{\rm hard}
 +{\cal F}_{2h}^{\rm hard}
\\
&\quad
 +{\cal F}_{2i}^{\rm hard}
 +{\cal F}_{2j}^{\rm hard}
\bigr],
\end{aligned}
\eeq
with individual contributions
\bqa
{\cal F}_{2a}^{\rm hard}
&=&
\left[-{5{\cal D}\over8}+{23\over32}\right]
I_{\rm ball}^{\rm bb},
\\
{\cal F}_{2b}^{\rm hard}
&=&
{1\over16}I_{\rm ball}^{\rm bb},
\\
{\cal F}_{2c}^{\rm hard}
&=&
{1\over32}I_{\rm ball}^{\rm bb},
\\
{\cal F}_{2d}^{\rm hard}
&=&
-{3\over16}{\cal D}({\cal D}-1)I_{\rm ball}^{\rm bb},
\\
{\cal F}_{2e}^{\rm hard}
&=&
{27\over16}({\cal D}-1)I_{\rm ball}^{\rm bb},
\eqa
\bqa
{\cal F}_{2f}^{\rm hard}
&=&
-{1\over4}
\left[
I_{\symod}^{\rm bb}
+I_{\symod}^{\rm bf}
+I_{\symod}^{\rm ff}
\right],
\\
{\cal F}_{2g}^{\rm hard}
&=&
{1\over8}I_{\rm ball}^{\rm bb},
\\
{\cal F}_{2h}^{\rm hard}
&=&
{{\cal D}-2\over8}\trI
\left[
{{\cal D}-6\over2}I_{\rm ball}^{\rm ff}
+(4-{\cal D})I_{\rm ball}^{\rm bf}
\right],
\nonumber\\
&&
\\
{\cal F}_{2i}^{\rm hard}
&=&
{({\cal D}-2)^2\over4}\trI
\left[
I_{\rm ball}^{\rm bf}
-2H_3
+f_2(f_1-b_1)^2
\right],
\nonumber\\
&&
\\
{\cal F}_{2j}^{\rm hard}
&=&
-{{\cal D}-2\over4}\trI\,I_{\rm ball}^{\rm bf}.
\eqa
The sum-integrals and the self-energy insertions
$I_{\symod}^{\rm bb}$, $I_{\symod}^{\rm bf}$, and $I_{\symod}^{\rm ff}$
are listed in Appendix~\ref{app:sumints} and
Appendix~\ref{app:hardalgebra}.
Two simplifications occur for the ${\cal D}_{\rm max}=6$ reduction.
First, the six contributions from diagrams with gauge and ghost lines
only, ${\cal F}_{2a}^{\rm hard}$--${\cal F}_{2e}^{\rm hard}$ and
${\cal F}_{2g}^{\rm hard}$, add up to
\beq
-{1\over16}({\cal D}-6)(3{\cal D}-2)\,I_{\rm ball}^{\rm bb},
\eeq
which vanishes identically at ${\cal D}=6$.  Second, the factor
$({\cal D}-6)$ in ${\cal F}_{2h}^{\rm hard}$ removes the
$I_{\rm ball}^{\rm ff}$ term.
Setting
\beq
  {\cal D}={\cal D}_{\rm max}=6,\qquad \trI=4
\eeq
gives the massless three-loop hard contribution for ${\cal N}=2$ SYM.  The
same algebra reproduces the known ${\cal N}=4$ result when ${\cal D}=10$ and
$\trI=8$.
Define
\beq
  {\cal B}\equiv {T^4\over144(4\pi)^2}.
\label{BunitN2}
\eeq
The three-loop hard contribution can then be written as
\beq
{\cal F}_{2}^{\rm hard}=d_A\lambda^2{\cal B}\,B_{\rm hard}^{(6)},
\eeq
where the superscript $(6)$ denotes the ${\cal D}_{\rm max}=6$ reduction,
not a loop order.
The coefficient is
\beq
\begin{aligned}
B_{\rm hard}^{(6)}
={}&
 -{54\over\epsilon}
 -324\Lb
 -288\Zone
\\
&
 +36\Zthree
 -72\gamma_E
 +{468\over5}\log2
 -252 .
\end{aligned}
\label{Bhard6N2}
\eeq
Normalizing the three-loop hard contribution to Eq.~\eqref{FidealN2} gives
\beq
\begin{aligned}
{{\cal F}_{2}^{\rm hard}\over{\cal F}_{\rm ideal}}
={}&
\x^2
\biggl[
{9\over32\epsilon}
{+}{27\over16}\Lb
{+}{3\over2}\Zone
\\
&\hspace{0.8cm}
-{3\over16}\Zthree
{+}{3\over8}\gamma_E
-{39\over80}\log2
{+}{21\over16}
\biggr].
\end{aligned}
\label{RhardN2}
\eeq
The pole in Eq.~\eqref{RhardN2} is canceled by the EFT unit-operator
counterterm together with the coupling-renormalization counterterm discussed
in Sec.~\ref{sec:eft}.

\subsection{Mass parameters}

\begin{figure*}[t]
\begin{center}
\begin{minipage}{0.48\textwidth}
\begin{center}
\includegraphics[width=0.96\linewidth]{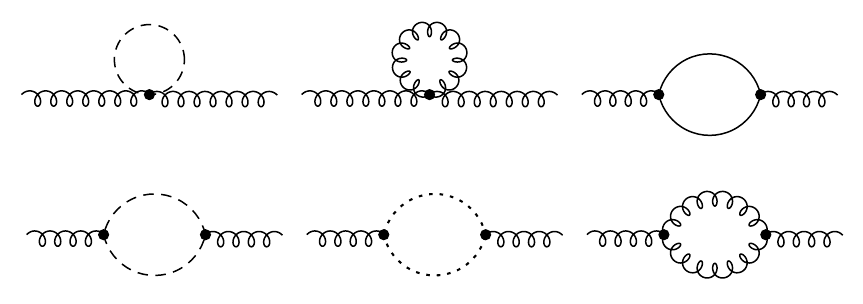}
\end{center}
\end{minipage}
\hfill
\begin{minipage}{0.48\textwidth}
\begin{center}
\includegraphics[width=0.96\linewidth]{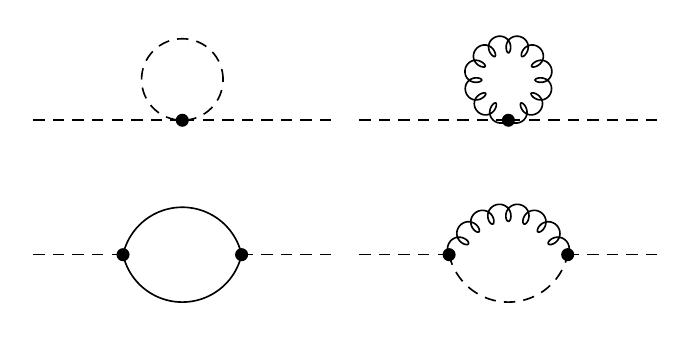}
\end{center}
\end{minipage}
\end{center}
\vspace{-4mm}
\caption{One-loop self-energy diagrams in the full four-dimensional
${\cal N}=2$ theory: gauge-field self-energy (left) and scalar self-energy
(right).  Spiral, dashed, solid, and dotted lines denote gauge fields,
adjoint scalars, adjoint fermions, and ghosts, respectively.  Reproduced
without modification from Ref.~\cite{Andersen:2022N4EFT}.}
\label{fig:fullSEN2}
\end{figure*}

\begin{figure*}[t]
\begin{center}
\begin{minipage}{0.48\textwidth}
\begin{center}
\includegraphics[width=0.95\linewidth]{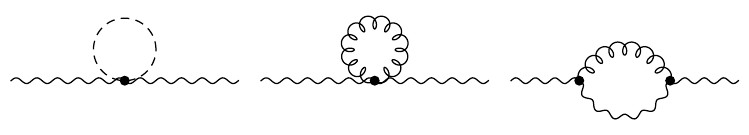}
\end{center}
\end{minipage}
\hfill
\begin{minipage}{0.48\textwidth}
\begin{center}
\includegraphics[width=0.95\linewidth]{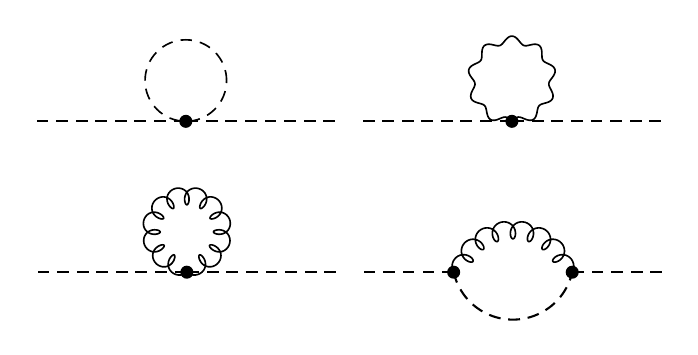}
\end{center}
\end{minipage}
\end{center}
\vspace{-4mm}
\caption{One-loop self-energies for $A_0$ (left) and $\Phi_I$ (right) in the
${\cal N}=2$ EFT.
Spiral lines denote three-dimensional gauge fields, sinusoidal lines denote
the adjoint scalar $A_0$, dashed lines denote the two real adjoint scalars
$\Phi_I$, and dotted lines denote ghosts.  Reproduced without modification
from Ref.~\cite{Andersen:2022N4EFT}.}
\label{fig:eftSEN2}
\end{figure*}

Matching the static two-point functions determines $m_E^2$ for $A_0$ and
$M^2$ for the two real adjoint scalars.  Their one-loop values provide the
hard contribution to the screening masses.  Beyond leading order, the
electric screening mass of a nonabelian gauge theory is infrared sensitive
and requires a nonperturbative definition~\cite{Arnold:1995bh}, although
its hard contribution remains computable order by order in strict
perturbation theory; only the one-loop values are needed here.
At leading order, the electric mass parameter is the static gauge-field
self-energy at zero momentum,
\beq
  m_E^2=\Pi_{00}(0,{\bf 0}),
\eeq
and the scalar mass parameter is the static scalar self-energy,
\beq
  M^2=\Sigma(0,{\bf 0}).
\eeq
The one-loop full-theory self-energy graphs are shown in
Fig.~\ref{fig:fullSEN2}.  The corresponding self-energy graphs in the
effective theory are shown in Fig.~\ref{fig:eftSEN2}; at leading order the EFT
integrals are scaleless in dimensional regularization, so the masses below
are fixed by the full-theory graphs.
Setting ${\cal D}_{\rm max}=6$ and $d=4$ in the one-loop matching formulas
gives the leading-order results
\bqa
  m_E^2&=&\lambda T^2+O(\lambda^2T^2),
\label{mEleadingN2}\\
  M^2&=&{\lambda T^2\over2}+O(\lambda^2T^2).
\label{MSleadingN2}
\eqa
For $SU(2)$, where $\lambda=2g^2$, these results give
$m_E^2=2g^2T^2+O(g^4T^2)$ and $M^2=g^2T^2+O(g^4T^2)$, in agreement with
the high-temperature masses of Paik and Yaffe and the leading $\symos$
result of Vazquez-Mozo~\cite{Paik:2009iz,VazquezMozo:1999ge}.
We keep the unexpanded RDR form of the one-loop matching coefficients for
the counterterm algebra.  The $\symod$ gauge-field self-energy gives
\bqa
m_E^2&=&
\lambda({\cal D}_{\rm max}-2)(d-2)(b_1-f_1),
\label{mEsymodN2}\\
M^2&=&
\lambda({\cal D}_{\rm max}-2)(b_1-f_1).
\label{MsymodN2}
\eqa
The first line follows from the static temporal component of the parent
gauge-field self-energy, while the second follows from an extra-dimensional
component of the same self-energy.
For $\symos$,
\beq
m_E^2=4\lambda(d-2)(b_1-f_1),
\qquad
M^2=4\lambda(b_1-f_1).
\eeq
Setting $d=4-2\epsilon$ in this form gives the $O(\epsilon)$ expansions.
The $O(\epsilon)$ part of these one-loop matching coefficients is needed
when it multiplies a $1/\epsilon$ soft pole.  In RDR,
\begin{align}
b_1-f_1
={}&{T^2\over8}\!
\biggl[
1+2\epsilon\biggl(
\log{\mub\over4\pi T}
+{\zeta'(-1)\over\zeta(-1)}
+1-\tfrac13\log2
\biggr)
\biggr].
\notag\\[-0.5ex]
&
\label{b1minusf1N2}
\end{align}
Thus
\begin{align}
m_E^2={}&\lambda T^2
\biggl[
1+\epsilon\biggl(
2\log{\mub\over4\pi T}
+2{\zeta'(-1)\over\zeta(-1)}
+1-{2\over3}\log2
\biggr)
\biggr],
\notag\\[-0.5ex]
&
\label{mEepsN2}
\end{align}
and
\begin{align}
M^2={}&{\lambda T^2\over2}
\biggl[
1+2\epsilon\biggl(
\log{\mub\over4\pi T}
+{\zeta'(-1)\over\zeta(-1)}
+1-\tfrac13\log2
\biggr)
\biggr].
\notag\\[-0.5ex]
&
\label{M2epsN2}
\end{align}

\subsection{Coupling constants}

Tree-level matching is sufficient for the effective couplings through
order $\lambda^2$.  Rescaling the static fields by $\sqrt{T}$, so that the
EFT fields have three-dimensional normalization, and comparing
$\int_0^\beta d\tau\,{\cal L}_{\symnt}$ with Eq.~\eqref{lagesymN2} gives
\bqa
  g_E^2&=&g^2T,\label{gE2N2}\\
  g_3^2&=&g^2T,\label{g32N2}\\
  h_E&=&g^2T.\label{hEN2}
\eqa
Unlike in ${\cal N}=4$ SYM, the four-dimensional gauge coupling of pure
${\cal N}=2$ SYM runs.  Its renormalization contributes to the free energy at
order $\lambda^2$.  Writing $g=g(\mub)$, where $\mub$ is the renormalization
scale, the one-loop beta function for $n_F$ adjoint Majorana fermions and
$n_S$ adjoint real scalars is
\beq
\begin{aligned}
  \mub{dg\over d\mub}&=-{b_0\over(4\pi)^2}g^3+O(g^5),
\\
  b_0&=\left({11\over3}-{2\over3}n_F-{1\over6}n_S\right)C_A .
\end{aligned}
\eeq
The three terms in $b_0$ are the gauge, fermion, and scalar contributions,
respectively.
For the ${\cal D}_{\rm max}=10$ reduction, $n_F=4$ and $n_S=6$, so
$b_0^{{\cal N}=4}=0$.  For ${\cal N}=2$ SYM, $n_F=2$ and $n_S=2$,
giving $b_0^{{\cal N}=2}=2C_A$~\cite{Seiberg:1988ur}.  In terms of
$\lambda=g^2C_A$,
\beq
\mub{d\lambda\over d\mub}
=-{\lambda^2\over4\pi^2}+O(\lambda^3),
\label{betaN2}
\eeq
or equivalently
\beq
\mub{d\x\over d\mub}
=-{1\over4}\x^2+O(\x^3).
\label{betaxN2}
\eeq
Here $\lambda\equiv\lambda(\mub)$ is the renormalized coupling.  The
dimensionful bare coupling is
\beq
\begin{aligned}
\lambda_B
&=\mub^{2\epsilon}\lambda
\left[
1-{\lambda\over8\pi^2\epsilon}+O(\lambda^2)
\right],
\\
\mub^{-2\epsilon}\lambda_B
&=\lambda\left[
1-{\lambda\over8\pi^2\epsilon}+O(\lambda^2)
\right].
\end{aligned}
\label{barelambdaN2}
\eeq
The second line is the dimensionless bare combination that multiplies the
sum-integral expressions.
At this order, the bare coupling is needed only for the coupling counterterm
in Sec.~\ref{sec:eft}.
For the temperature-dependent plots in Sec.~\ref{sec:result} we use the
one-loop solution $\lambda(\mub)/\pi^2=4/\log(\mub/\Lambda)$ at
$\mub=4\pi T$, where $\Lambda$ is the RG-invariant scale.
We choose the hard--soft factorization scale equal to the
renormalization scale $\mub$.  The factorization-scale
dependence cancels between $f_E$ and $f_M$; the residual $\mub$ dependence of
the final result is fixed by Eq.~\eqref{betaN2}.

\section{Soft contributions and counterterms}
\label{sec:eft}

The soft contribution is obtained from the EFT with the matched parameters
of Sec.~\ref{sec:params}.  The $A_0$ and $\Phi_I$ propagators contain the
masses $m_E$ and $M$, whereas the magnetostatic gauge field remains massless.
Together with the hard term $Tf_E$ and the counterterms below, the following
diagrams complete the free energy through order $\lambda^2$.
The one-loop graphs contributing to the soft free energy are shown in
Fig.~\ref{fig:soft1N2}.  Evaluating the diagrams using the integrals in
Appendix~\ref{app:3dints} gives
\begin{figure}[t]
\begin{center}
\includegraphics[width=0.95\linewidth]{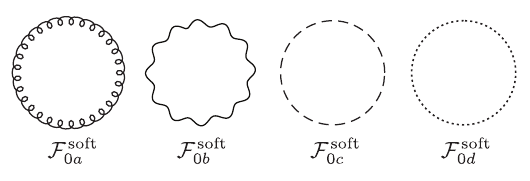}
\end{center}
\vspace{-4mm}
\caption{One-loop soft vacuum diagrams in the ${\cal N}=2$ EFT.  The magnetostatic gauge
and ghost loops are scaleless in
dimensional regularization and do not contribute to Eq.~\eqref{FM1N2}; only
the massive $A_0$ and $\Phi_I$ loops survive.  Line conventions follow
Fig.~\ref{fig:eftSEN2}.  Reproduced without modification from
Ref.~\cite{Andersen:2022N4EFT}.}
\label{fig:soft1N2}
\end{figure}
\beq
\begin{aligned}
f_{M,1}
&=
-{1\over2}d_A
\left[
I_0'(m_E^2)+2I_0'(M^2)
\right]
\\
&=
-{d_A\over12\pi}\left[m_E^3+2M^3\right].
\end{aligned}
\label{FM1N2}
\eeq
For ${\cal N}=2$ SYM this gives the standard EFT ring
contribution~\cite{Braaten:1995jr,Andersen:2022N4EFT}, with the
${\cal N}=4$ coefficient $m_E^3+6m_S^3$ replaced by $m_E^3+2M^3$.  The
four-dimensional contribution is $T f_{M,1}$, which gives the
order-$\lambda^{3/2}$ term
\beq
{T f_{M,1}\over{\cal F}_{\rm ideal}}
=
\left(1+{1\over\sqrt2}\right)
\left({\lambda\over\pi^2}\right)^{3/2}.
\label{lambda32N2}
\eeq

\subsection{Two-loop soft contribution}

\begin{figure*}[t]
\begin{center}
\includegraphics[width=0.65\textwidth]{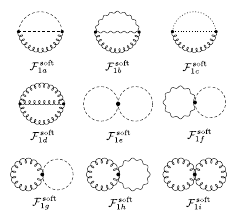}
\end{center}
\vspace{-4mm}
\caption{Two-loop soft vacuum diagrams in the ${\cal N}=2$ EFT.  Line conventions follow
Fig.~\ref{fig:eftSEN2}.  Reproduced without modification from
Ref.~\cite{Andersen:2022N4EFT}.}
\label{fig:soft2N2}
\end{figure*}

The two-loop graphs contributing to the soft free energy are shown in
Fig.~\ref{fig:soft2N2}.  The same sunset and double-bubble topologies occur in
the ${\cal N}=4$ EFT calculation.
The six real adjoint scalars of ${\cal N}=4$ are replaced by
two real adjoint scalars.
For a theory with $n_s$ real adjoint scalars, all with mass $M$, the
two-loop EFT expression before inserting the explicit integrals is
\beq
\begin{aligned}
T f_{M,2}={}&
T d_A C_A\biggl\{
g_E^2\left[
{1\over4}I_1^2(m_E^2)
+m_E^2J_1(m_E^2)
\right]
\\
&
 +n_s g_E^2\left[
{1\over4}I_1^2(M^2)
+M^2J_1(M^2)
\right]
\\
&
 +{n_s\over2}h_E I_1(m_E^2)I_1(M^2)
 +{n_s(n_s-1)\over4}g_3^2 I_1^2(M^2)
\biggr\}.
\end{aligned}
\label{genericsoft2ns}
\eeq
Diagrams containing only massless magnetostatic gauge and ghost propagators
are scaleless and vanish in dimensional regularization.
For $n_s=6$ this reproduces the ${\cal N}=4$ coefficients
$6$, $3$, and $15/2$ multiplying the scalar-gauge, mixed, and scalar-quartic
terms, respectively.  For ${\cal N}=2$ SYM one sets $n_s=2$, giving
coefficients $2$, $1$, and $1/2$.
We have explicitly checked that the one- and two-loop contributions are
independent of the covariant-gauge parameter $\xi$, as shown in
Appendix~\ref{app:xi}.
The ultraviolet pole of $J_1(m^2)$ is removed by the EFT unit-operator
counterterm
\beq
\delta f_E^{{\cal N}=2}
=
-{d_A C_A\over4(4\pi)^2\epsilon}\,g_E^2
\left[m_E^2+2M^2\right].
\label{deltafEN2}
\eeq
Adding this counterterm gives the finite two-loop EFT contribution, with
the factorization scale kept explicit,
{\small
\beq
\begin{alignedat}{1}
T\left(f_{M,2}+\delta f_E^{{\cal N}=2}\right)
={}&
\tfrac{\dA\lambda T^2}{(4\pi)^2}
\!\biggl[
m_E^2\left(\tfrac34+\log\tfrac{\mub}{2m_E}\right)
\\
&
+2M^2\left(\tfrac34+\log\tfrac{\mub}{2M}\right)
 +m_EM+{1\over2}M^2
\biggr].
\end{alignedat}
\label{Fsoft2N2}
\eeq
}
The last two terms are the mixed $A_0\Phi_I$ double-bubble and the scalar
commutator-quartic contribution, respectively.  The mixed term is derived in
Appendix~\ref{app:ctalg}.
Using the leading-order masses in Eqs.~\eqref{mEleadingN2} and
\eqref{MSleadingN2},
\beq
\begin{gathered}
T\left(f_{M,2}+\delta f_E^{{\cal N}=2}\right)
={}
{\dA\lambda^2T^4\over16\pi^2}
\biggl[
{7\over4}+{1\over\sqrt2}+{5\over2}\log2
\\
\quad
+2\Lb-\log{\lambda\over\pi^2}
\biggr].
\end{gathered}
\eeq
Writing ${\cal F}_{\rm soft}=T[f_{M,1}+f_{M,2}+\delta f_E^{{\cal N}=2}]$,
the normalized soft contribution through order $\lambda^2$ is
\beq
\begin{aligned}
{{\cal F}_{\rm soft}\over{\cal F}_{\rm ideal}}
={}&
\left(1+{1\over\sqrt2}\right)
\x^{3/2}
+\x^2
\biggl[
{3\over4}\log\x
\\
&\hspace{0.8cm}
-{3\over2}\Lb
-{21\over16}
-{3\sqrt2\over8}
-{15\over8}\log2
\biggr].
\end{aligned}
\label{RsoftN2}
\eeq
The first term in Eq.~\eqref{RsoftN2} is the ring, or plasmon,
contribution and agrees with the high-temperature result of Paik and Yaffe
after setting the gauge group to $SU(2)$.  The same normalized coefficient
follows from the general ring result of
Vazquez-Mozo~\cite{Paik:2009iz,VazquezMozo:1999ge}.  The order-$\lambda^2$
soft term is the two-loop EFT contribution computed here and combines with
the hard contribution and counterterms below.

\subsection{Unit-operator counterterm in the hard matching}

For $SU(N_c)$, the counterterm in Eq.~\eqref{deltafEN2} follows from the ${\cal N}=4$
counterterm by replacing $6m_S^2$ with
$2M^2$~\cite{Andersen:2022N4EFT}.  Since the mass parameters multiply the
$1/\epsilon$ pole, their RDR $O(\epsilon)$ parts must be retained.  In the
hard matching, the corresponding term is $-T\delta f_E^{{\cal N}=2}$.  In
units of ${\cal B}$ defined in Eq.~\eqref{BunitN2}, expanding
$-T\delta f_E^{{\cal N}=2}$ with the RDR mass parameters gives the
scalar and gluon pieces listed in Appendix~\ref{app:ctalg}:
% labeled by the subscripts ${\rm s.ct}$ and ${\rm g.ct}$
\begin{align}
B_{\rm s.ct}^{(6)}
&=
{36\over\epsilon}
+72\Lb
+72\Zone
-24\log2
\notag\\
&\quad
+72,
\label{BscalarctN2}\\
B_{\rm g.ct}^{(6)}
&=
{36\over\epsilon}
+72\Lb
+72\Zone
-24\log2
\notag\\
&\quad
+36 .
\label{BgluonctN2}
\end{align}
The different finite constants come from the $O(\epsilon)$ parts of the
one-loop matching coefficients for $m_E^2$ and $M^2$.
Their sum, the unit-operator counterterm contribution, is
% (subscript ${\rm u.ct}$)
\beq
\begin{aligned}
B_{\rm u.ct}^{(6)}
={}&
 {72\over\epsilon}
 +144\Lb
 +144\Zone
 -48\log2
\\
&
 +108 .
\end{aligned}
\label{BthermalctN2}
\eeq
Equivalently,
\beq
\begin{aligned}
{-T\delta f_E^{{\cal N}=2}\over{\cal F}_{\rm ideal}}
={}&
\x^2
\biggl[
-{3\over8\epsilon}
-{3\over4}\Lb
\\
&\hspace{0.8cm}
-{3\over4}\Zone
{+}{1\over4}\log2
-{9\over16}
\biggr].
\end{aligned}
\label{RthermalctN2}
\eeq
Combining Eqs.~\eqref{RhardN2} and \eqref{RthermalctN2} gives
\beq
\begin{aligned}
&{{\cal F}_{2}^{\rm hard}-T\delta f_E^{{\cal N}=2}\over{\cal F}_{\rm ideal}}
\\
&\quad=
\x^2
\biggl[
-{3\over32\epsilon}
{+}{15\over16}\Lb
{+}{3\over4}\Zone
\\
&\quad\hspace{0.8cm}
-{3\over16}\Zthree
{+}{3\over8}\gamma_E
-{19\over80}\log2
{+}{3\over4}
\biggr].
\end{aligned}
\label{RhardthermalN2}
\eeq
\subsection{Coupling-renormalization counterterm}

The pole structure differs from that of ${\cal N}=4$.  In the conformal
${\cal N}=4$ theory the corresponding
EFT unit-operator counterterm cancels the hard pole without any coupling
renormalization.  Here it leaves a residual pole, which is removed by the
coupling-renormalization counterterm.

The coupling-renormalization counterterm is obtained by substituting the
dimensionless bare combination in Eq.~\eqref{barelambdaN2} into the
order-$\lambda$ hard contribution.  We write
\beq
{{\cal F}_{1}^{\rm hard}\over{\cal F}_{\rm ideal}}
=
\x\left[r_0+\epsilon r_1\right],
\qquad
r_0=-{3\over4},
\eeq
with $r_1$ read from Eq.~\eqref{OepslambdaN2}.  The second line of
Eq.~\eqref{barelambdaN2} gives
\[
{\mub^{-2\epsilon}\lambda_B\over\pi^2}
=\x\left[1-{\x\over8\epsilon}\right],
\]
and hence
\beq
{\delta_\lambda{\cal F}_{1}^{\rm hard}\over{\cal F}_{\rm ideal}}
=
-{\x^2\over8\epsilon}
\left[r_0+\epsilon r_1\right].
\label{deltalambdaoriginN2}
\eeq
Thus the coupling-renormalization contribution is
\beq
\begin{aligned}
{\delta_\lambda{\cal F}_{1}^{\rm hard}\over{\cal F}_{\rm ideal}}
={}&
\x^2
\biggl[
{3\over32\epsilon}
{+}{3\over8}\Lb
\\
&\hspace{0.8cm}
{+}{3\over8}\Zone
-{1\over8}\log2
{+}{3\over8}
\biggr].
\end{aligned}
\label{RcouplingctN2}
\eeq
Adding this to Eq.~\eqref{RhardthermalN2} cancels the remaining pole,
\beq
  -{3\over32\epsilon}+{3\over32\epsilon}=0.
\eeq
The finite order-$\lambda^2$ hard-plus-counterterm contribution is therefore
\beq
\begin{aligned}
&{{\cal F}_{2}^{\rm hard}-T\delta f_E^{{\cal N}=2}+\delta_\lambda{\cal F}_{1}^{\rm hard}\over{\cal F}_{\rm ideal}}
\\
&\quad=
\x^2
\biggl[
{21\over16}\Lb
{+}{9\over8}\Zone
-{3\over16}\Zthree
\\
&\quad\hspace{0.8cm}
{+}{3\over8}\gamma_E
-{29\over80}\log2
{+}{9\over8}
\biggr].
\end{aligned}
\label{RhardctN2}
\eeq

\section{Final result and checks}
\label{sec:result}

\begin{figure*}[t]
\begin{center}
\includegraphics[width=0.95\textwidth]{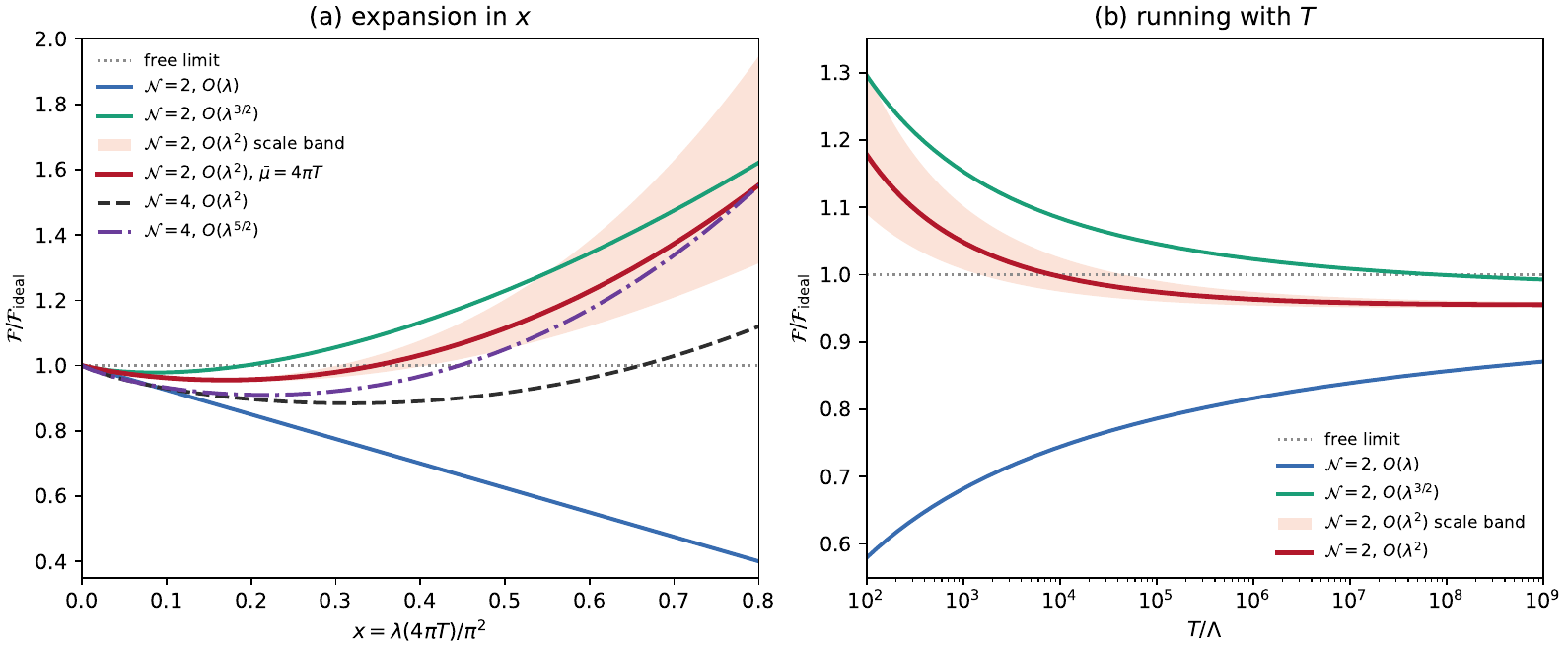}
\end{center}
\caption{Normalized pure ${\cal N}=2$ SYM free energy.  Panel (a) shows
successive weak-coupling truncations as a function of
$x=\lambda(4\pi T)/\pi^2$;
the dashed and dot-dashed curves show the conformal ${\cal N}=4$ SYM
results through orders $\lambda^2$ and
$\lambda^{5/2}$, respectively~\cite{Andersen:2022N4EFT,Carrington:2026N4lambda52}.
Panel (b) shows the corresponding ${\cal N}=2$ truncations as a function of
$T/\Lambda$ using the one-loop running
$\lambda(4\pi T)/\pi^2=4/\log(4\pi T/\Lambda)$.  In both panels the
shaded band shows the order-$\lambda^2$ variation
$\bar\mu/(4\pi T)\in[1/2,2]$.  The larger-$x$ region illustrates the
separation of successive truncations and is not a claim of perturbative
accuracy.}
\label{fig:n2convergence}
\end{figure*}

Adding the hard and soft parts gives
\beq
\begin{aligned}
\frac{\mathcal{F}}{\mathcal{F}_{\rm ideal}^{\mathcal{N}=2}}
={}&
1-{3\over4}\x
+\left(1+{1\over\sqrt2}\right)\x^{3/2}
+\x^2
\\
&
\biggl[
{3\over4}\log\x
-{3\over16}\Lb
-{3\over16}
-{3\sqrt2\over8}
{+}{3\over8}\gamma_E
\\
&\hspace{-0.15cm}
{+}{9\over8}\Zone
-{3\over16}\Zthree
-{179\over80}\log2
\biggr]
+O(\lambda^{5/2}).
\end{aligned}
\label{finalN2}
\eeq
This is the normalized free energy of the high-temperature phase through
order $\lambda^2$ for $SU(\Nc)$ at arbitrary $\Nc$, with
$\lambda=g^2\Nc$ and $d_A=\Nc^2-1$.
The behavior of successive weak-coupling truncations is shown in
Fig.~\ref{fig:n2convergence}, both as a function of
$x=\lambda(4\pi T)/\pi^2$ and,
using the one-loop running coupling, as a function of $T/\Lambda$.

\subsection{Comparison with Paik and Yaffe}

Paik and Yaffe quote the high-temperature equilibrium free-energy density
of pure $SU(2)$ ${\cal N}=2$ SYM as~\cite{Paik:2009iz}
\beq
{\cal F}
=
3T^4\left[
-{\pi^2\over12}
+{g^2\over8}
-{1+\sqrt2\over6\pi}g^3
+O(g^4)
\right].
\label{PaikYaffeN2}
\eeq
For $SU(2)$, $C_A=2$, $d_A=3$, and $\lambda=2g^2$.  Using
Eq.~\eqref{finalN2} through order $g^3$ gives
\bqa
{\cal F}
&=&
-{3\pi^2T^4\over12}
\biggl[
1-{3\over4}{2g^2\over\pi^2}
\nonumber\\
&&\hspace{0.7cm}
+\left(1+{1\over\sqrt2}\right)
\left({2g^2\over\pi^2}\right)^{3/2}
\biggr]
+O(g^4)
\nonumber\\
&=&
3T^4\left[
-{\pi^2\over12}
+{g^2\over8}
-{1+\sqrt2\over6\pi}g^3
\right]
+O(g^4),
\eqa
in exact agreement.  The result of Ref.~\cite{Paik:2009iz} was obtained for
$SU(2)$ only, so this comparison is necessarily limited to $\Nc=2$.

\subsection{RG check}
\label{sec:rgcheck}

The explicit scale dependence in Eq.~\eqref{finalN2} is fixed by
Eq.~\eqref{betaxN2}.  The running of the order-$\lambda$ term gives
\beq
\left(-{1\over4}\right)\left(-{3\over4}\right)
\left({\lambda\over\pi^2}\right)^2
=
{3\over16}\left({\lambda\over\pi^2}\right)^2.
\eeq
This is canceled by the derivative of the explicit scale logarithm
$-(3/16)(\lambda/\pi^2)^2\log(\mub/4\pi T)$.  Thus
\beq
\left(-{1\over4}\right)\left(-{3\over4}\right)-{3\over16}=0,
\eeq
as required by RG invariance through order $(\lambda/\pi^2)^2$.

\subsection{Pole cancellation}

The pole remaining after adding the EFT unit-operator counterterm to the hard
contribution is
\beq
\left[
{{\cal F}_{2}^{\rm hard}-T\delta f_E^{{\cal N}=2}\over{\cal F}_{\rm ideal}}
\right]_{\rm pole}
=
-{3\over32\epsilon}\x^2,
\eeq
while the coupling-renormalization counterterm gives
\beq
\left[
{\delta_\lambda{\cal F}_{1}^{\rm hard}\over{\cal F}_{\rm ideal}}
\right]_{\rm pole}
=
{3\over32\epsilon}\x^2.
\eeq
The total order-$\lambda^2$ result is therefore finite.

\subsection{\texorpdfstring{${\cal N}=4$}{N=4} hard-sector check}

We checked the hard $\symod$ algebra in Appendix~\ref{app:hardalgebra}
by setting ${\cal D}_{\rm max}=10$ and $\trI=8$.  Adding the corresponding
${\cal N}=4$ EFT unit-operator counterterm cancels the hard pole without
coupling renormalization, as in the conformal ${\cal N}=4$
calculation~\cite{Andersen:2022N4EFT}.  This checks the
dimensional-reduction algebra before setting ${\cal D}_{\rm max}=6$.

\section{Summary and outlook}
\label{sec:conclusions}

We computed the finite-temperature free energy density of pure
four-dimensional ${\cal N}=2$ SYM through order $\lambda^2$.  The hard
contribution is obtained by evaluating the massless $\symos$ vacuum diagrams
through three loops with loop momenta in $d=4-2\epsilon$ dimensions.  The
soft contribution is computed in the three-dimensional EFT obtained by
high-temperature dimensional reduction.  Relative to the conformal
${\cal N}=4$ calculation, the running of the ${\cal N}=2$ gauge coupling
leaves a residual pole after the EFT unit-operator counterterm;
this pole is canceled by the coupling-renormalization counterterm.  The
resulting renormalization-scale dependence is fixed by the one-loop beta
function.

The separation of hard and soft contributions accounts for the logarithmic
structure.  The $\log(\lambda/\pi^2)$ term comes from the soft scale, as in the conformal
${\cal N}=4$ calculation, while the explicit
$\log(\bar\mu/4\pi T)$ dependence is fixed by the one-loop beta function of
${\cal N}=2$ SYM.  At order $\lambda^2$, no higher powers of logarithms
occur, consistent with the standard dimensional-reduction structure of
QCD~\cite{Braaten:1995jr}.

The order-$\lambda^2$ contribution computed here has not, to our knowledge,
been obtained previously; prior work extended the high-temperature pure
${\cal N}=2$ SYM free energy through the $O(g^3)$ ring
contribution~\cite{Paik:2009iz,VazquezMozo:1999ge}.  The result can be
compared with future lattice studies of
finite-temperature pure ${\cal N}=2$ SYM, or extended perturbatively to
${\cal N}=2$ theories with hypermultiplet matter.  The next perturbative
term is the order-$\lambda^{5/2}$ correction.  As in QCD and ${\cal N}=4$
SYM, this term is perturbative and arises entirely from the soft scale.  It
requires the three-loop massive vacuum diagrams of the effective theory and
the next-to-leading matching of the mass parameters.  At order $\lambda^3$,
the magnetic scale enters, and purely perturbative calculations are no
longer sufficient~\cite{Linde:1980ts,Gross:1980br}.

\section*{Acknowledgements}

The work of Q.D.\ is supported by the Guangxi Natural Science Foundation
under Grant No.~2023GXNSFBA026027.  U.T.\ acknowledges support from the
Department of Mathematics and Natural Sciences at Prince Mohammad Bin Fahd
University.
We thank Jens O. Andersen and Michael Strickland for their collaboration on
the earlier ${\cal N}=4$ effective-field-theory calculation, whose hard/soft
setup is used here.

\appendix

\section{Sum-integrals}
\label{app:sumints}

The bosonic and fermionic sum-integrals are defined as
\bqa
\sumint_P&=&
\left({e^{\gamma_E}\mub^2\over4\pi}\right)^\epsilon
T\sum_{P_0=2\pi nT}
\nonumber\\
&&\times
\int{d^{3-2\epsilon}p\over(2\pi)^{3-2\epsilon}},
\\
\sumint_{\{P\}}&=&
\left({e^{\gamma_E}\mub^2\over4\pi}\right)^\epsilon
T\sum_{P_0=(2n+1)\pi T}
\nonumber\\
&&\times
\int{d^{3-2\epsilon}p\over(2\pi)^{3-2\epsilon}}.
\eqa
We use
\beq
  b_n=\sumint_P{1\over P^{2n}},\qquad
  f_n=\sumint_{\{P\}}{1\over P^{2n}},\qquad n\geq0.
\eeq
The one-loop sum-integrals needed are
\bqa
b_0'&=&{\pi^2T^4\over45}\left[1+O(\epsilon)\right],
\label{b0primeN2}\\
f_0'&=&-{7\pi^2T^4\over360}\left[1+O(\epsilon)\right],
\label{f0primeN2}
\eqa
\bqa
b_1&=&{T^2\over12}
\biggl[
1+\epsilon\biggl(
2\Lb+2\Zone+2
\biggr)
\biggr]+O(\epsilon^2),
\nonumber\\[-0.5ex]
&&\label{b1N2}\\
f_1&=&-{T^2\over24}
\biggl[
1+\epsilon\biggl(
2\Lb
\nonumber\\
&&\hspace{0.8cm}
+2\Zone+2-2\log2
\biggr)
\biggr]+O(\epsilon^2),
\label{f1N2}\\
b_2&=&{1\over(4\pi)^2}
\left[
{1\over\epsilon}+2\Lb+2\gamma_E
\right]+O(\epsilon),
\label{b2N2}\\
f_2&=&{1\over(4\pi)^2}
\biggl[
{1\over\epsilon}+2\Lb+2\gamma_E+4\log2
\biggr]+O(\epsilon).
\nonumber\\[-0.5ex]
&&\label{f2N2}
\eqa
Primes on $b_0$ and $f_0$ denote differentiation with respect to their
propagator exponent.
The two-loop massless sum-integrals which appear in intermediate
three-loop reductions vanish in dimensional regularization~\cite{Arnold:1994ps}:
\bqa
\sumint_{PQ}{1\over P^2Q^2(P+Q)^2}&=&0,
\\
\sumint_{\{P\}Q}{1\over P^2Q^2(P+Q)^2}&=&0,
\\
\sumint_{\{PQ\}}{1\over P^2Q^2(P+Q)^2}&=&0.
\eqa
For compactness, we express the results below in terms of the unit
\beq
  {\cal B}=
  {1\over(4\pi)^2}\left({T^2\over12}\right)^2,
\eeq
introduced in Eq.~\eqref{BunitN2}.
The three-loop basketball and auxiliary sum-integrals are, through order
\(\epsilon^0\),
\begin{align}
I_{\rm ball}^{\rm bb}
={}&
{\cal B}\biggl[
{6\over\epsilon}
+36\Lb
-12\Zthree
\nonumber\\
&\hspace{0.65cm}
+48\Zone
+{182\over5}
\biggr],
\label{IbbN2}\displaybreak[2]\\
I_{\rm ball}^{\rm ff}
={}&
{\cal B}\biggl[
{3\over2\epsilon}
+9\Lb
-3\Zthree
\nonumber\\
&\hspace{0.65cm}
+12\Zone
+{173\over20}
-{63\over5}\log2
\biggr],
\label{IffN2}\displaybreak[2]\\
I_{\rm ball}^{\rm bf}
={}&
-{1-2^{11-3d}\over6}
I_{\rm ball}^{\rm bb}
-{1\over6}I_{\rm ball}^{\rm ff},
\label{IbfN2}\displaybreak[2]\\
H_3
={}&
{\cal B}\biggl[
{3\over8\epsilon}
+{9\over4}\Lb
+{3\over2}\Zthree
-{3\over2}\Zone
\nonumber\\
&\hspace{0.65cm}
+{9\over4}\gamma_E
+{361\over160}
+{57\over10}\log2
\biggr],
\label{H3N2}\displaybreak[2]\\
H_4
={}&
{\cal B}\biggl[
{5\over24\epsilon}
+{5\over4}\Lb
-{1\over6}\Zthree
+{7\over6}\Zone
\nonumber\\
&\hspace{0.65cm}
+{1\over4}\gamma_E
+{23\over24}
-{8\over5}\log2
\biggr],
\label{H4N2}\displaybreak[2]\\
H_5
={}&
{\cal B}\biggl[
{4\over3\epsilon}
+8\Lb
-{5\over3}\Zthree
\nonumber\\
&\hspace{0.65cm}
+{26\over3}\Zone
+\gamma_E
+{49\over12}
\biggr],
\label{H5N2}\displaybreak[2]\\
H_6
={}&
{\cal B}\biggl[
-{17\over48\epsilon}
-{17\over8}\Lb
+{5\over24}\Zthree
-{11\over6}\Zone
\nonumber\\
&\hspace{0.65cm}
-{1\over2}\gamma_E
-{41\over48}
+{11\over8}\log2
\biggr].
\label{H6N2}
\end{align}
The basketball sum-integrals and $H_3$ were calculated in
Refs.~\cite{Arnold:1994ps,Arnold:1994eb}.  The remaining auxiliary
sum-integrals were calculated in Ref.~\cite{Du:2021pqa}.

\section{Integrals in the effective theory}
\label{app:3dints}

The three-dimensional integrals are defined by
\beq
\int_p=
\left({e^{\gamma_E}\mub^2\over4\pi}\right)^\epsilon
\int{d^{3-2\epsilon}p\over(2\pi)^{3-2\epsilon}}.
\eeq
We use
\beq
I_n(m^2)=\int_p{1\over(p^2+m^2)^n}.
\eeq
The prime on $I_0'(m^2)$ denotes differentiation with respect to the
exponent $n$.
The two-loop sunset integrals are
\beq
J_n(m^2)=
\int_{pq}
{1\over
(p^2+m^2)(q^2+m^2)^n(p-q)^2}.
\eeq
The integrals needed through two loops are
\bqa
I_0'(m^2)
&=&
{m^3\over4\pi}
\left({\mub\over2m}\right)^{2\epsilon}
\left[
{2\over3}+{16\over9}\epsilon+O(\epsilon^2)
\right],
\\
I_1(m^2)
&=&
{m\over4\pi}
\left({\mub\over2m}\right)^{2\epsilon}
\left[-1-2\epsilon+O(\epsilon^2)\right],
\\
J_1(m^2)
&=&
{1\over(4\pi)^2}
\left({\mub\over2m}\right)^{4\epsilon}
\left[
{1\over4\epsilon}+{1\over2}+O(\epsilon)
\right].
\eqa

\section{Hard \texorpdfstring{$\symod$}{SYM(1,D)} algebra}
\label{app:hardalgebra}

This appendix lists the $\symod$ self-energy combinations that enter
the main-text three-loop vacuum diagrams, as obtained in
Refs.~\cite{Du:2021pqa,Andersen:2022N4EFT}.  They are
\bqa
I_{\symod}^{\rm bb}
&=&
{({\cal D}-2)^2\over4}\bar I_{\symod}^{\rm bb}+2{\cal D}I_{\rm ball}^{\rm bb},
\\
I_{\symod}^{\rm ff}
&=&
{(\trI)^2\over4}
\left[
\bar I_{\symod}^{\rm ff}
+({\cal D}-3)I_{\rm ball}^{\rm ff}
\right],
\\
I_{\symod}^{\rm bf}
&=&
-\trI
\left[
{{\cal D}-2\over2}\bar I_{\symod}^{\rm bf}
+{3\over2}({\cal D}-2)I_{\rm ball}^{\rm bf}
\right],
\nonumber\\
&&
\eqa
with
\bqa
\bar I_{\symod}^{\rm bb}
&=&
4({\cal D}-4)b_2b_1^2
+16H_5-I_{\rm ball}^{\rm bb},
\\
\bar I_{\symod}^{\rm ff}
&=&
4({\cal D}-4)b_2f_1^2
+16H_4-I_{\rm ball}^{\rm ff},
\\
\bar I_{\symod}^{\rm bf}
&=&
4({\cal D}-4)b_2b_1f_1
+16H_6-I_{\rm ball}^{\rm bf}.
\eqa
Direct substitution of ${\cal D}=6$ and $\trI=4$ gives
Eq.~\eqref{Bhard6N2}.

\section{Counterterm and soft algebra}
\label{app:ctalg}

This appendix collects counterterm and soft-sector algebra specific to
${\cal N}=2$ SYM.

\subsection{Mass parameters through order \texorpdfstring{$\epsilon$}{epsilon}}

The bosonic and fermionic one-loop sum-integrals are related by
\beq
\begin{aligned}
f_1&=\left(2^{-1+2\epsilon}-1\right)b_1
\\
&=\left(-{1\over2}+\epsilon\log2\right)b_1+O(\epsilon^2),
\end{aligned}
\eeq
with $b_1$ given in Eq.~\eqref{b1N2}.  Therefore
\beq
b_1-f_1
=
{3\over2}b_1-\epsilon(\log2)\,b_1+O(\epsilon^2),
\eeq
which yields Eq.~\eqref{b1minusf1N2}.
For $\symos$, the unexpanded RDR matching gives
$m_E^2=4\lambda(d-2)(b_1-f_1)$ and
$M^2=4\lambda(b_1-f_1)$.  These self-energy matching identities give
Eqs.~\eqref{mEepsN2} and \eqref{M2epsN2}.

\subsection{Mixed soft double-bubble}

The mixed $A_0\Phi_I$ double-bubble follows directly from the EFT operator
$h_E{\rm Tr}[(i[A_0,\Phi_I])^2]$.  With $n_s=2$ real adjoint scalars,
\bqa
T f_{M,2}^{A_0\Phi}
&=&
T\,d_A C_A\,{n_s\over2}h_E I_1(m_E^2)I_1(M^2)
\nonumber\\
&=&
T\,d_A C_A\,h_E
{m_E M\over(4\pi)^2}
={d_A\lambda T^2\over(4\pi)^2}m_E M,
\eqa
where $h_E=g^2T$ and $\lambda=g^2C_A$.  This gives the coefficient of the
$m_E M$ term in Eq.~\eqref{Fsoft2N2}.

\subsection{Expansion of the EFT unit-operator counterterm}

We expand the scalar and gluon terms in $-T\delta f_E^{{\cal N}=2}$ through
order $\epsilon^0$.  The $O(\epsilon)$ parts of the RDR mass parameters must
be retained because they multiply the pole in Eq.~\eqref{deltafEN2}.  The
mass ratios are
\beq
{M^2\over\lambda T^2}
={1\over2}
\left[
1+\epsilon\left(
2\Lb+2\Zone+2-{2\over3}\log2
\right)
\right],
\eeq
\beq
{m_E^2\over\lambda T^2}
=
1+\epsilon\left(
2\Lb+2\Zone+1-{2\over3}\log2
\right).
\eeq
In units of ${\cal B}=T^4/[144(4\pi)^2]$, the scalar and gluon terms are
\bqa
B_{\rm s.ct}^{(6)}
&=&
144\,{2M^2\over\lambda T^2}
\left({1\over4\epsilon}\right),
\label{BsctformulaN2}\\
B_{\rm g.ct}^{(6)}
&=&
144\,{m_E^2\over\lambda T^2}
\left({1\over4\epsilon}\right).
\label{BgctformulaN2}
\eqa
Expanding Eqs.~\eqref{BsctformulaN2} and \eqref{BgctformulaN2} through
order $\epsilon^0$ reproduces Eqs.~\eqref{BscalarctN2} and
\eqref{BgluonctN2}.  The coupling-renormalization counterterm is derived in
Sec.~\ref{sec:eft}.

\subsection{Gauge-parameter independence}
\label{app:xi}

In general covariant gauge, the propagator of the gauge field in the effective
theory depends on $\xi$ only through a term proportional to
$k_ik_j/(k^2)^2$, where $k$ is the momentum of the gauge field.  At one loop,
the massive $A_0$ and $\Phi_I$ loops do not involve this propagator, and the
massless gauge-field and ghost loops vanish in dimensional regularization for
any $\xi$.  At two loops, diagrams containing only massless propagators also
vanish, and so do the double-bubble diagrams ${\cal F}_{1g}^{\rm soft}$ and
${\cal F}_{1h}^{\rm soft}$ in Fig.~\ref{fig:soft2N2}, which contain a massless
gauge-field loop.  Of the remaining diagrams, only the sunset diagrams
${\cal F}_{1a}^{\rm soft}$ and ${\cal F}_{1b}^{\rm soft}$ involve the
gauge-field propagator.  Let $p$ and $q$ be the momenta of the massive
propagators.  The gauge-field momentum is $k=p-q$, and the momentum factor at
each vertex is $(p+q)_i$.  Since $(p+q)\cdot k=p^2-q^2$, the $\xi$-dependent
term is proportional to
\beq
\begin{aligned}
&\int_{pq}{(p^2-q^2)^2\over(k^2)^2(p^2+m^2)(q^2+m^2)}
\\
&\quad=2\int_{pq}{p^2-q^2\over(k^2)^2(q^2+m^2)}
\\
&\quad=2\int_{kq}{2q\cdot k+k^2\over(k^2)^2(q^2+m^2)}
=2I_1(m^2)\int_k{1\over k^2}=0,
\end{aligned}
\label{xiproofN2}
\eeq
where $m=m_E$ for $A_0$ and $m=M$ for $\Phi_I$.  In the first step, we used
$p^2-q^2=(p^2+m^2)-(q^2+m^2)$ and the symmetry $p\leftrightarrow q$.  In the
second step, we changed variables from $p$ to $k$.  The term proportional to
$q\cdot k$ is odd in $k$ and integrates to zero.  The remaining massless tadpole
is scaleless and vanishes in dimensional regularization.  Thus
Eqs.~\eqref{FM1N2} and \eqref{genericsoft2ns} are independent of $\xi$.
\par\vfill

\balance

\end{document}